# Peroperative Shear-Wave Elastography for the Evaluation of Spinal Cord Release in Pediatric Dysraphism: First Experience in 21 Patients

Mathilde Gaumé [1*] MD, Lou Richard [1] MD, Claudio Vergari [2] PhD,
Eleonore Blondiaux [3] MD PhD, Marius Buffard [1] MD,
Raphael Vialle [1] MD PHD, Timothée de Saint Denis[1] MD.

M. Gaumé and L. Richard contributed the same.

## Abstract

**Objectives**: Shear wave elastography (SWE) is a recent method for evaluating soft tissue stiffness. Its application to the spinal cord is limited by the ultrasound bone barrier. Spinal dysraphisms can cause neurological symptoms impairing bladder, bowel and lower limb due to a mechanical stress known as "tethered cord syndrome." Detethering of the spinal cord is a part of the surgical treatment. This study aims to assess the feasibility of intraoperative elastography to guide surgery in patients with spinal dysraphism involving traction of the spinal cord.

**Methods**: This single-center retrospective study included all patients who underwent surgery for spinal dysraphism between November 2023 and July 2024. SWE was performed under general anesthesia with the patient in the prone position before and after spinal cord detethering. All SWE measurements were performed using the Aixplorer® system. Demographic, clinical and radiological data (age at surgery, neurological status, MRI findings) were collected.

**Results:** 21 patients with an average age of 3.7 years ± 5.3 (range, 0-17.5) were included. Etiologies included open saccular dysraphism (n=3), spinal cord conus lipoma (n=6), Limited Dorsal Myeloschisis (n=3) and fibrolipomatous filum (n=9). In 85% of patients, at least one clip was of sufficient quality for analysis both before and after release. The mean spinal cord stiffness at the beginning of surgery was 25.2 ± 9.2 kPa (range, 12.3-39.1) and 11.2 ± 3.8 kPa (range, 6.1-16.5) at the end of release, with a significant stiffness decrease ($p=0.018$). No complications attributable to SWE were observed.

**Conclusion**: SWE of the spinal cord in patients with spinal dysraphism is a safe and feasible imaging method. This technique may assist in assessing the adequacy of untethering during surgery. However, no established normative SWE values for the pediatric spinal cord currently exist. Until it does, the intraoperative interpretability and clinical utility of this technique remain limited.

**Keywords:** Shear Wave Elastography, Spinal dysraphism, Tethered cord release, Spinal cord stiffness, Intraoperative imaging

1 University Institute for Spine Surgery, Armand Trousseau Hospital, Sorbonne University, Paris, France.

2 Arts et Metiers Institute of Technology, EPF Engineering School, Université Sorbonne Paris Nord, IBHGC–Institut de Biomécanique Humaine Georges Charpak, F-75013 Paris, France.

3 Department of Pediatric Radiology, Trousseau Hospital, APHP Sorbonne Université, Paris, France.

* Corresponding author: Mathilde Gaume. mathilde.gaume@aphp.fr

## Introduction

Spinal dysraphism can be classified according to the extent of mechanical stress on the spinal cord1,2, primarily resulting from traction, as seen in tethered cord syndrome3, that can lead to urinary, bowel or neuro-orthopedic impairment 4,5. Surgical treatment, which requires a canal opening, is guided by balancing the neurological risk of natural history against surgical risk. The procedure aims to optimize the spinal cord environment affected by malformation, with the release of spinal cord tethering being a key part.

However, one of the major challenges encountered in routine practice is the lack of reliable exams to assess neurological risk and surgical potential benefits. Additionally, intraoperatively, surgeons do not have a tool to guide their detethering procedure beyond the improvement of neuromonitoring data when these are compromised preoperatively. Indeed, in severe malformative cases such as chaotic lipomas complete detethering may not be achieved due to short and malformative spinal nerve roots. Incorporating intraoperative measurements could help to reach an optimized mechanical loading of the spinal cord. Spinal ultrasound is performed for the diagnostic within the first month of life 6. However, its reliability decreases over time due to ossification of the posterior vertebral arch7.

Elastography allows objective and reproducible assessment of tissue stiffness 8, 9. In the most recent elastography techniques, stiffness is estimated by measuring the speed of shear waves through the tissue 10,11. A color-coded elastogram superimposed on B-mode imaging is directly visualized, enabling the operator to be guided by both anatomical structures and tissue stiffness when selecting a region of interest. It is now commonly used for liver fibrosis and breast cancer 12, and is currently being extensively studied for other indications, particularly as a tool in the operating room 13-15.

It is gradually gaining popularity in neurosurgery, a field where manual palpation is discouraged by tissue fragility. Bone anechogenicity is a major obstacle for the central nervous system assessment, requiring craniectomy or laminectomy for this evaluation15. Most studies focus on the use of elastography for brain tumor resection, with fewer studies addressing epilepsy and hydrocephalus 14,15. Only three studies 16-18 have investigated its use on human spinal cord, with one focusing on compressed spinal cord cases and the other two on conditions with tethered cord syndrome in adult patients.

The primary objective of this study was to assess the feasibility of SWE at the spinal cord level in operating room for pediatric patients with spinal dysraphism. The secondary objective was to evaluate if the stiffness of the spinal cord in patients undergoing surgery for dysraphism decreases between the beginning of the procedure and after release of tethering decompression. We also aimed to provide an average stiffness measurement at the beginning and end of the procedure for different types of pathology (conus lipoma, myelomeningocele, Myé Limited Dorsal Myeloschisis (MyéLDM), open saccular, tract dysraphism and fibrolipomatous filum.

## Methods

This retrospective study reviewed all consecutive patients with spinal dysraphism who underwent surgery between November 2023 and July 2024. All procedures were conducted at our referral pediatric spinal surgery center by a single experienced surgeon.

Data collected included demographic information (age, sex, etiology, presence of a cutaneous and/or neurological abnormalities), and SWE measurement. The study was approved by the appropriate ethics committee (IRB approval Nr FHUMAMUTH20025-13). Informed consent for the study and the publication of the data was obtained from the parents/guardians of all participants.

All SWE measurements were performed using the Aixplorer® system (Supersonic Imagine, Aix-en-Provence, France). It utilizes 2D SWE, and measurements are expressed as elastic modulus in kilopascals (kPa). Measurements were obtained from 30-second clips of a sagittal median view of the spinal cord to ensure reproducibility. It has been demonstrated that shear wave velocities vary depending on whether the slices are axial or sagittal (shear wave speed is faster in sagittal and slower in axial views) [19]. All measurements were performed directly on the machine by selecting regions of interest (ROI) in the "tracing" mode, to allow maximum flexibility in selecting the area for analysis. The tracing included the pia mater, the spinal cord, and the central canal (Figure 1).

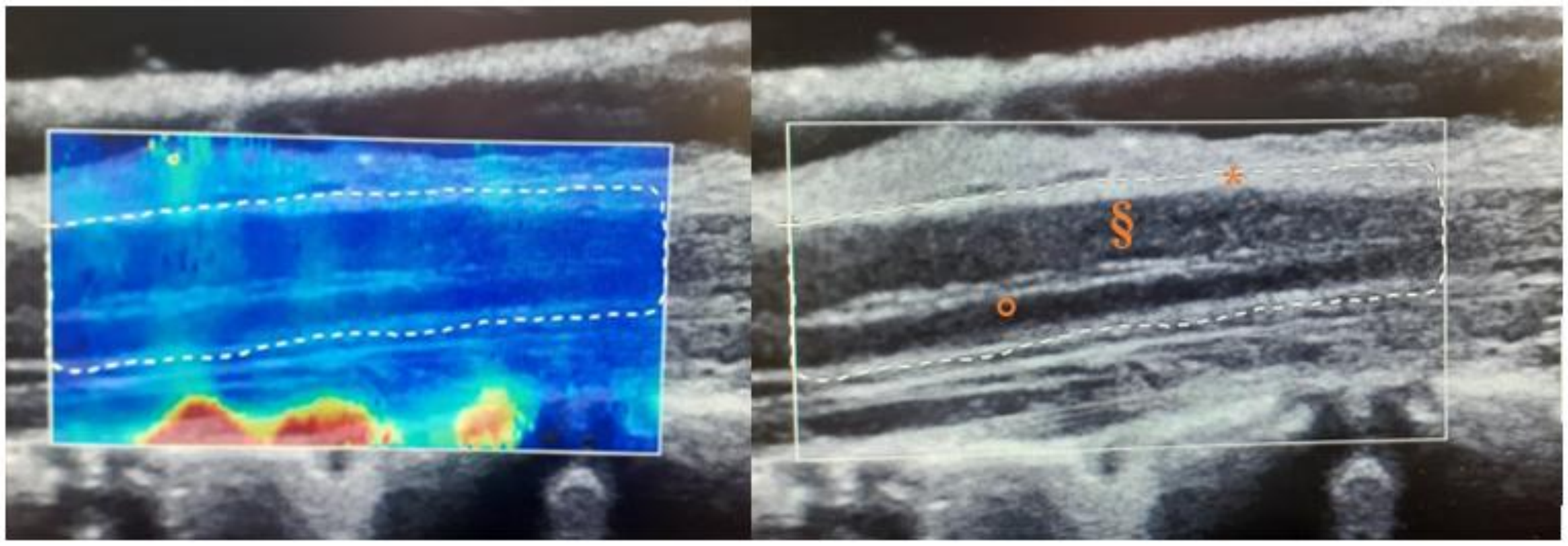

**Figure 1.** Example of spinal cord elastography with ROI selection. (A) Elastographic image with ROI selection (---). (B) Corresponding B-mode ultrasound image, including the pia mater (*), spinal cord tracts (§), and the central canal (°).

The ROI tracing was performed on three different images selected from the 30-second clip. The data displayed represents an average of all values within the region of interest. It is presented with its mean, maximum and minimum values, and standard deviation (SD).

## Intraoperative protocol

All acquisitions were performed under general anesthesia with the patient in the prone position. An initial SWE was made after patient positioning and prior skin preparation.

The 10–2 MHz linear probe (with a ~4 cm measurement surface) was used when the extent of the laminectomy allowed it, while the 20–6 MHz "golf-club" linear probe (with a ~3 cm measurement surface) was used when the laminectomy was shorter. Sterile ultrasound gel was applied to the closed end of a sterile tubular cover in which the probe was placed (CIV-Flex™ Transducer Cover). The assembly was secured to the sterile field using adhesive tape or a surgical clamp, leaving enough length for free manipulation.

The surgical site was filled with sterile, body-temperature water to the brim (to avoid air bubble artifacts), and the covered probe was submerged contacting the spinal cord, ensuring no compressive force was applied.

Ideally, acquisitions were performed during four surgical stages after laminotomy:

Before releasing the spinal cord tethering: before and/or after opening the dura mater

After releasing the spinal cord tethering: before and/or after closing the dura mater

Continuous monitoring of hemodynamic parameters was conducted, along with neurological monitoring through motor and somatosensory evoked potentials (MEP/SSEP).

## Statistical analysis

Descriptive variables are expressed as mean and standard deviation. The normality of the variables was tested using the Shapiro-Wilk test. The primary outcome, evaluating the difference in stiffness before and after release, and the subgroup analyses were assessed using the Wilcoxon test. The comparison of spinal cord stiffness according to the presence or absence of a neurological disorder was performed using the Mann-Whitney U test. All statistical analyses were conducted using SPSS (version 26, IBM).

## Results

### Demographic data

Over a period of 9 months, elastography measurements were performed on 21 patients in the operating room. The main demographic data are summarized in Table 1.

A terminal detethering with filum section was associated in 5 cases: 3 dysraphisms with tract, and 2 meningoceles. It is important to highlight that for one meningocele, the filum was described as normal during surgery.

**Table 1**: Demographic characteristics of patients who underwent surgery for spinal dysraphism

| | MyeLDM | Tract Dysraphism | Conus lipoma | Meningocele | Filum | TOTAL |
|---|---|---|---|---|---|---|
| Number of patients, | 3 | 3 | 6 | 2 | 7 | 21 |
| Mean age | 1 day | 6.6 years | 3.5 years | 4.2 months | 5.5 years | 3.7 years |
| Sex (F/M) | 2/1 | 0/3 | 5/1 | 1/1 | 5/2 | 13/8 |
| Preoperative Neurological deficit | 1 (Club Foot) | 0 | 4 (Bladder dysfunction ± neurogenic bowel and/or lower limb deficit, equinus) | 0 | 2 (Varus foot, constipation) | 7 |
| Cutaneous anomaly | 3 | 3 | 6 | 2 | 2 | 16 |
| Antenatal diagnosis | 3 | 0 | 2 | 2 | 0 | 7 |

### Feasibility of measurements

Regarding the feasibility of intraoperative elastography of the 21 patients, 2 had no high-quality clips available for elastography measurements (one tract dysraphism and one fibrolipomatous filum), accounting for 10% of the patients. This may be due to technical issues with the machine, intraoperative technical difficulties that did not allow for an elastography phase, or a laminectomy that was too small relative to the size of the probe. One patient had only one clip of sufficient quality for measurements (one thoracic tract), representing 5% of the patients.

The remaining patients (except fibrolipomatous patients) had at least one clip at the beginning of the procedure and one at the end to allow for high-quality measurements (85% of the patients), patients with an isolated abnormal filum had only one clip of the filum taken before sectioning due to the minimal invasive procedure without spinal cord exposure.

## Spinal cord stiffness at the beginning and at the end of the procedure

The stiffness values at the beginning and end of the procedure by type of malformation are summarized in Table 2.

Across all types of malformations (excluding filums), the overall stiffness at the beginning of the procedure before opening of the dura mater was 25.2 ± 9.2 kPa (min 12.3, max 39.1), and at the end of the procedure before closure of the dura mater, it was 11.2 ± 3.8 kPa (min 6.2, max 16.5). The average reduction was 13.9 ± 6.8 kPa, corresponding to a decrease of 54% ± 13%. The stiffness of the spinal cord decreased significantly ($p = 0.018$). (Figure 2 and 3).

For statistical analysis, paired elastography clip measurements were available in only seven patients, with analyzable data acquired prior to dural opening and prior dural closure.

The average filum stiffness was 11.7±7.1kPa, ranging from 1.6 to 24.8kPa.

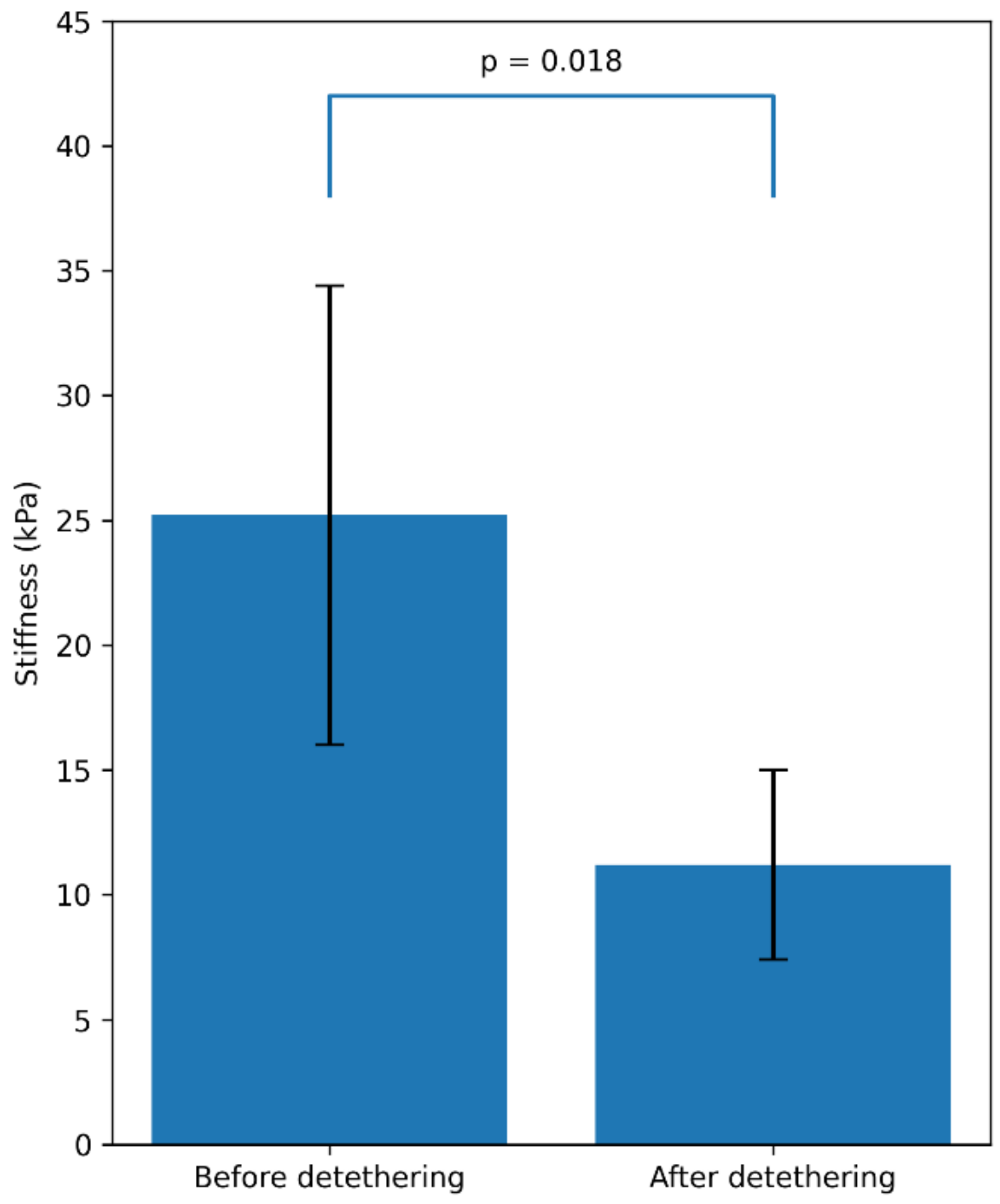

**Figure 2**. Evolution of spinal cord stiffness in dysraphism between the beginning and the end of the procedure. Data are presented as mean ± standard deviation. *p = 0.018.

## Stiffness according to the opening of the dura mater

For 2 dysraphisms - 1 conus lipoma (n°3 in table 2) and the filum close to the medullary (n°15), measurements were taken before and after the opening of the dura mater. The stiffness before the opening of the dura mater was 15.4±3.6 kPa (min 12.1, max 19.5). The average stiffness of the spinal cord after the opening of the dura mater was 9.9±1.4 kPa (min 7.1, max 11.7).

For four dysraphisms - 3 conus lipoma (n°1, 3, 6) and the filum close to the medullary (n°15) measurements were taken before and after the closure of the dura mater. The average stiffness of the spinal cord before the closure of the dura mater was 8.7±2.3 kPa (min 5.4, max 12.6). The average stiffness of the spinal cord after the closure of the dura mater was 11.75±6.8 kPa (min 3.5, max 25).

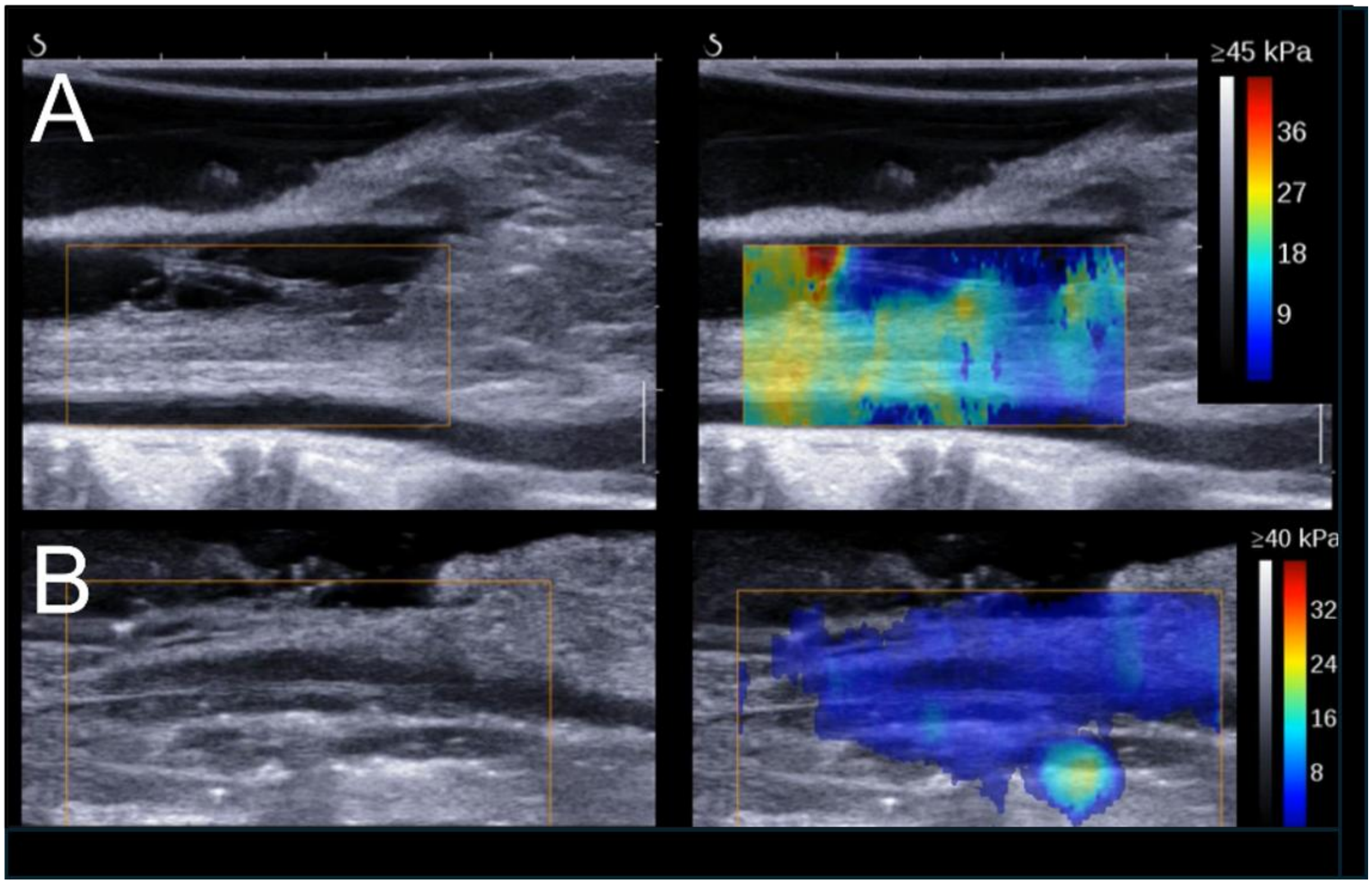

**Figure 3.** Elastography images of the spinal cord in a patient undergoing surgery for a conus medullaris lipoma. (A) At the beginning of the procedure, longitudinal ultrasound and shear-wave elastography demonstrate a globally increased mean spinal cord stiffness. (B) At the end of the procedure, the stiffness map shows a marked decrease in spinal cord stiffness, reflecting reduced mechanical constraint following surgical decompression. Color bars indicate stiffness values in kPa.

## Post operative complications

Four cases of postoperative neurological complications were observed. This number of complications is similar to the expected complication rate for this type of procedure, with no evidence suggesting it is secondary to the use of elastography.Two occurred in patients with conus lipoma: one presented with a left lower limb motor deficit with partial resolution, and the other with persistent urinary retention requiring hetero catheterization. Two additional cases were reported in patients who underwent sectioning of the filum: one experienced transient urinary urgency for one month, and the other developed temporary hypoesthesia of the anterior thigh and dorsum of the left foot at three weeks postoperatively.

One case of postoperative hydrocephalus occurred following the closure of an open spinal dysraphism at birth, without clinical impact and with no need for shunt placement. One early postoperative infection was observed in a patient with conus lipoma, caused by *E. coli* and *Enterococcus Gallinarum,* and was successfully treated with surgical lavage and intravenous antibiotic therapy.

**Table 2**: Evolution of spinal cord stiffness between the beginning, before and after opening of the dura mater, and end of the procedure, before and after closure of the dura mater, for each patient (except other filum ).

| | Stiffness at the beginning of the procedure (kPa) Mean ± SD (min - max) | | Stiffness at the end of the procedure (kPa) Mean ± SD (min - max) | |
|---|---|---|---|---|
| Patient number and etiology | Before opening the dura mater | After opening the dura mater | Before closure of the dura mater | After closure of the dura mater |
| 1: Conus lipoma | 21.8 ± 0.4 (21.4-22.2) | / | 11.0 ± 1.5 (9.8-12.6) | 10.1 ± 3.3 (8.0-13.9) |
| 2: Conus lipoma | 26.2 ± 0.6 (25.8-26.9) | / | / | / |
| 3: Conus lipoma | 18.5 ± 1.7 (16.6-19.5) | 10.1 ± 2.6 (7.1-11.7) | 6.2 ± 1.1 (5.4-7.4) | 14.4 ± 1.4 (13.5-16.0) |
| 4: Conus lipoma | 32.3 ± 3.0 (29.0-34.7) | / | 13.5 ± 0.2(13.3-13.7) | / |
| 5: Conus lipoma | / | 10.0 ± 0.5 (9.5-10.5) | 7.8 ± 1.9(6.1-9.8) | / |
| 6: Conus lipoma | 30.4 ± 2.0 (28.5-32.4) | / | 10.6 ± 0.8(9.8-11.4) | 13.4 ± 1.4(12.4-15) |
| 7: MyeLDM | 26.4 ± 5.2(23.1-32.4) | / | / | 9.9 ± 1.4 (8.5-11.3) |
| 8: MyeLDM | 34.6 ± 2.3 (32.5 – 39.1) | / | 13.7 ± 2.5 (9.9- 16.5) | / |
| 9: MyeLDM | 24.4 ± 2.3(19.2-25.1) | / | / | 14.0 ± 2.3(11.4-15.8) |
| 10 : Tract dysraphism | / | / | / | / |
| 11 : Tract dysraphism | 26.6 ± 3.3(16.7-22.5) | / | 13.2 ± 0.9(12.1-13.9) | / |
| 12 : Tract dysraphism | 23.6 ± 2.2(21.4-25.7) | / | / | / |
| 13: Meningocele | / | / | / | / |
| 14 : Meningocele | / | 10.95 ± 0.7(8.8-10.2) | / | 7.9 ± 1.4(7.5-9.9) |
| 15: Filum | 12.3±0.6(12.3-12.6) | 9.6±0.7(8.8-10.2) | 9.8±1.2 (6.2-10.1) | 9.1±1.4(7.5-9.9) |

## Discussion

In the present series, we demonstrated that the stiffness of the spinal cord in patients undergoing surgery for dysraphisms decreased between the beginning and the end of the procedure.

To our knowledge, this is the most notable series to show actual clinical applicability to spinal pathology. The feasibility demonstrated in this series supports the potential role of SWE as an adjunct to intraoperative decision-making. In cases such as conus lipomas or complex associated malformations, visual assessment of residual tethering may be challenging. In these situations, SWE could provide an additional quantitative parameter reflecting residual spinal cord stiffness. For example, if the considered baseline is reached, the detethering procedure can be considered achieved and may avoid increasing risk with a more extensive surgery. If the considered baseline is not reached and the surgical detethering appears optimized (with only residual short nerve roots), a complementary procedure can be proposed.

Only three previous studies [16-18] have investigated spinal cord stiffness, but in adult patients, and their results were consistent with ours. While baseline values were heterogeneous, the end-of-procedure measurements were more consistent, highlighting their potential utility for intraoperative monitoring.

Compared to the earliest study on compressed spinal cords [16], which reported a mean stiffness of 93.84 kPa with values varying according to the cause of compression, our measurements at the beginning of the procedure were lower (likely due to differences in patient populations and underlying pathologies). After decompression, their reported stiffness values were 11.05 kPa for the pia mater, 9.35 kPa for the spinal cord tracts, and 9.95 kPa for the central canal. Methodologically, they used three circular ROIs of 1 mm in diameter placed on specific structures (dura mater, pia mater, spinal cord tracts, and central canal), whereas we selected a single ROI encompassing the entire visible spinal cord excluding the dura mater. As in our results, stiffness values were more variable in compressed cords and more consistent following decompression. Such changes in variabilities could be expected since preoperatively the spinal cord is subjected to different subject-specific loads, while postoperatively the spinal cord should be closer to its physiological loading.

The second study was a case report by the same team [17]. They investigated spinal cord stiffness in a 45-year-old patient presenting a tethered cord syndrome, in a degenerative context of high-grade L5/S1 spondylolisthesis with a complex malformative background consisting in a split cord malformation at L3, an intradural spinal cord lipoma, and a dermal sinus at L3/L4. Their results were very similar to ours, as stiffness of the terminal cone at the beginning of the procedure was 20-24 kPa, and after decompression was < 11 kPa.

The most recent study investigated spinal cord stiffness under traction using finite element simulation on anatomical models and in 6 adults who underwent spinal column shortening for tethered cord syndrome [18]. They also demonstrated a decrease in spinal cord stiffness from the beginning to the end of the procedure. However, the equipment they used differed from ours, which makes direct comparison of the results more difficult. Moreover, they report their results in m/s, but conversion can be done using the following formula: speed = √(elastic modulus / (3 * 1.1)).

Indeed, given the high variability of techniques and devices, caution is needed when comparing results. Similarly, differences in measurements may arise between studies due to the use of different probes with varying frequencies. It is important to note that we used the same elastography machine as in the studies by El-Habib [16, 17, 20].

One of the main questions is the influence of the dural opening versus the physiological watertight dural sheath in SWE intraoperative data. At the beginning of the procedure, the non-negligible decrease in stiffness (-5.5 kPa) after the dura mater opening might be explained by the significant loss of cerebrospinal fluid, leading to a reduction in the perimedullary environment to atmospheric pressure levels. At the end of the procedure, the dura mater was closed with a running suture of 5/0 or 6/0 absorbable monofilament. Before complete closure, Ringer Lactate fluid was infused into the intradural space to remove air and blood, and restore intradural pressure, thereby preventing epidural venous bleeding. The re-pressurization within the dura mater was confirmed by an increase in stiffness (3.2 kPa). However, the reliability of SWE after dura mater closure was highly variable, affecting both image quality and measurements due to persistent air bubbles and the compressive effect of the saline. In the case study reported by Almotairi et al. [17], measurements before release were taken after laminectomy and before the dura mater was opened, while measurements at the end of the procedure were performed before the dura mater was closed. We did not find any in vivo studies in the literature comparing spinal cord elasticity before and after dura mater opening.

## SWE to support intraoperative decision-making

As a new tool in the surgeon's belt, SWE could enhance both efficiency and safety in numerous applications. In neurosurgery, elastography has been primarily investigated for its ability to guide surgical resection of tumor lesions, detect residual tumor presence , or identify lesions that may be invisible to the surgeon and conventional imaging techniques [20-24].

In the context of dysraphism, the primary surgical objective is to achieve effective untethering. B-mode ultrasound could, similarly to what Pang demonstrated with MRI, help assess a favorable sac-cord ratio [25,] and elastography could be an additional tool to guide surgical decision-making (complementing neuromonitoring and visual assessment). Its quantitative nature may offer more precision in evaluating the extent of release. However, larger studies with more homogeneous patient groups are needed to validate its significance and clinical relevance.

## Limitations

Several limitations were identified in our study.

First, the number of patients was small, due to the low incidence of the pathology, and the ethical requirements of a feasibility study.

An open question remains as to whether the post-therapeutic stiffness values correspond to a "normal" baseline. Currently, no reference stiffness metrics are available to define normal values or to confirm SWE normalization in tethered cord syndrome. Therefore, postoperative SWE values, particularly those in the higher range observed after detethering, cannot be interpreted as definitive indicators of complete mechanical release. Larger prospective studies including correlation with postoperative clinical outcomes will be required to prevent a false-negative measurement after detethering.

One team used magnetic resonance elastography (MRE), a technique also based on the propagation of shear waves[26], to study spinal cord stiffness in healthy volunteers [27], as cited in El Habib's study [16]. The reported spinal cord stiffness was 12.1 kPa. Other studies on the biomechanical properties of the spinal cord were conducted on animal models or ex vivo [28, 29, 30], which do not provide accurate measurements, as they excluded factors that contributed to tissue stiffness in vivo, such as vascularization [31].

In 15% of patients, at least one clip was uninterpretable, most often due to technical issues. The main technical challenge was probe sizes, which were sometimes unsuitable for younger patients relative to the size of the laminotomy. In such cases, the "golf club" probe had to be used. However, its lower penetration occasionally resulted in absent SWE signals in the B-mode image.

Measurement differences between studies may also arise from the use of probes with different frequencies. However, in two studies [16,17] on the human spinal cord, the same machine as ours was employed.

Finally, we did not assess inter-operator variability. We had considered performing multi-operator elastography clips in the operating room, but it was not feasible to avoid extending the surgical duration and raise ethical concerns.

## Conclusions

Intraoperative SWE was feasible in the operating room, after surgical exposure and opening of the spinal canal. To our knowledge, this is the first study to explore spinal cord SWE in the pediatric population. This technique could be a useful tool for surgeons to determine if the spinal cord is sufficiently untethered, without prolonging operative time. Dysraphic spinal cord lipomas appeared to be particularly suitable for SWE, as their stepwise surgical management allows elastographic confirmation of surgical outcomes. However, no established normative SWE values for the pediatric spinal cord currently exist. Until it does, the intraoperative interpretability and clinical utility of this technique will remain limited.

**Data Access Statement**: The data presented in this study are available on request from the corresponding author. The data are not publicly available due to privacy concerns.

**Ethical Compliance:**

IRB approval Nr FHUMAMUTH20025-13. All the patients of the series gave informed consent for the study and the publication of the data.